\documentclass[letterpaper]{article} 
\usepackage{aaai23}  
\usepackage{times}  
\usepackage{helvet}  
\usepackage{courier}  
\usepackage[hyphens]{url}  
\usepackage{graphicx} 
\usepackage{natbib}  
\usepackage{caption} 
\usepackage{algorithm}
\usepackage{algorithmic}

\usepackage{newfloat}
\usepackage{listings}
\DeclareCaptionStyle{ruled}{labelfont=normalfont,labelsep=colon,strut=off} 
\floatstyle{ruled}
\newfloat{listing}{tb}{lst}{}
\floatname{listing}{Listing}
\title{How Do US Congress Members Advertise Climate Change: \\ An Analysis Of Ads Run On Meta's Platforms}
\author {
    Laurenz Aisenpreis\textsuperscript{\rm 1,*},
    Gustav Gyrst\textsuperscript{\rm 1,*},
    Vedran Sekara\textsuperscript{\rm 1,$\dagger$}
}
\affiliations {
    \textsuperscript{\rm 1} IT University of Copenhagen \\
    la.aisenpreis@gmail.com, ggyrst@gmail.com, vsek@itu.dk \\
    \textsuperscript{$\dagger$} These authors contributed equally \\
    \textsuperscript{$\dagger$} Corresponding author
}

\begin{document}

\maketitle

\begin{abstract}
Ensuring transparency and integrity in political communication on climate change has arguably never been more important than today.
Yet we know little about how politicians focus on, talk about, and portray climate change on social media.
Here we study it from the perspective of political advertisement.
We use Meta's Ad Library to collect 602,546 ads that have been issued by US Congress members since mid-2018.
Out of those only 19,176 (3.2\%) are climate-related.
Analyzing this data, we find that Democrats focus substantially more on climate change than Republicans, with 99.7\% of all climate-related ads stemming from Democrats.
In particular, we find this is driven by a small core of Democrats, where 72\% of all impressions can be attributed to 10 politicians.
Interestingly, we find a significant difference in the average amount of impressions generated per dollar between the two parties.
Republicans generate on average 188\% more impressions with their climate ads for the same money spent as Democrats.
We build models to explain the differences and find that demographic factors only partially explain the variance.
Our results demonstrate differences of climate-related advertisements of US congress members and reveal differences in advertising characteristics between the two political parties.
We anticipate our work to be a starting point for further studies about climate-related ads on Meta's platforms.
\end{abstract}

\section{Introduction} \label{introduction}

Climate change is considered one of the biggest, if not the, greatest challenge of our time~\cite{IPCC2021, climate_change_health, UN_climate_change_biggest_threat}.
Despite the large scientific consensus about the causes of climate change~\cite{IPCC2021}, it is still considered a complex policy issue, without any agreement on how to address it~\cite{victor2015climate}.

Social media has in the last two decades been used to communicate political messages, and is today considered an integral part of the political toolbox~\cite{newmediapolitics}.
For instance, social media has been found to have been instrumental in both Barack Obama's and Donald Trump's presidential campaigns~\cite{cogburn2011networked,allcott2017social}. 
It has changed traditional political activities and enabled political actors to: publicly express their opinion, reach different and broader networks~\cite{tarai2015political, nott2020political}, engage with their audience in new ways~\cite{kearney2017interpersonal}, and raise funds for campaigns~\cite{auter2018social}.
In addition, social media has also changed how political campaigns are conducted, allowing politicians to micro-target potential voters and frame their messaging strategically~\cite{sahly2019social}. \citet{fowler2021political} found that US politicians' Facebook ads occur earlier in campaigns, are less negative, less issue-focused, and more partisan than television advertising.

Today, parties and politicians spend extensive amounts of money on social media campaigns.
Political online advertising spending has more than quadrupled from 2018 to 2020, and political actors spent \$\emph{2.3 billion} on Meta and Google alone in 2019 and 2020~\cite{tech4campaigns}.
Unlike traditional means of political communication, social media is not a `one-to-many' channel, but rather a two-sided `many-to-many' communication channel. 
On the one hand, political actors can respond to the priorities of potential voters, evaluate engagement, and embed feedback and insights from online campaigns to optimize their political strategies~\cite{ennser2022does}.
On the other hand, recipients of political communication can be influenced, and potentially be informed by the political content they are exposed to~\cite{bode2016political}.
In spite of the relevancy of social media, we still know little about how political communication on climate change is characterized on social media.
Existing literature focuses on social media discussions of climate change, and less on the political discourse around it and advertisement about it~\cite{williams2015network, pearce2019social, sarewitz2011does}.
As such, there is a gap in research on how politicians advertise climate change on social media. 

Here we focus on Meta's platforms (Facebook, Messenger, Instagram, WhatsApp), with Facebook being the largest social media platform in terms of users~\cite{statista-facebook}.
Facebook is the preferred channel for people to get their news from, for instance, a third of American adults receive their news through the platform~\cite{matsa2021news}.
We leverage Meta's Ad Library\footnote{The API documentation can be found at: \url{https://www.facebook.com/ads/library/api/}.} data to gain insights into the advertisement activities of Congress members, and specifically to understand how they speak about climate change.
In particular, we focus on analyzing relevant metrics such as spend, impressions, geographic, and demographic coverage of their climate and non-climate related ads.
We focus on US Congress members for two reasons: 1) as the world's leading economy, the United States, will play a vital role in addressing climate change, and 2) Congress is the legislative branch of the US, with the power to shape future policies for the country.

The rest of the paper is organized as follows. 
First, we outline the state-of-the-art research within the field. Second, we describe the process of collecting our comprehensive dataset of ads for Congress members. Third, we summarize our findings, and lastly, we discuss limitations and possible extensions of our work.

\section{Related literature}
Social media is an important medium for spreading news and shaping opinions, and thus an interesting avenue for researchers to examine the gap between the scientific, public, and political opinions on climate change.
The existing literature on climate change predominantly focuses on discussions on Twitter, which can be divided into three main categories: public (e.g. users' knowledge and views on the topic), themes (e.g. studies based on thematic data sets such as hashtags or keywords), and professional communications (e.g. communication of climate scientists on social media)~\cite{pearce2019social}.
\citet{cody2015climate} measure sentiment to analyze how Twitter users respond to different climate change news, events, and natural disasters.
They find that climate change topics related to natural disasters, climate bills, and oil-drilling decrease happiness, while topics such as climate rallies increase happiness.
Further, they find that Twitter serves as a valuable tool to spread awareness of climate change, and that the voice of activists has a larger presence online than skeptics and deniers~\cite{cody2015climate}.
\citet{williams2015network} find there is high polarization on the topic of climate change among
users on Twitter, where users with strong opinions (i.e. skeptics and activists) are the most vocal. 
When it comes to US politicians talking about climate change, \citet{yu2021tweeting} find that the likelihood of a Democrat tweeting about the topic is associated with the existing public opinion, not the climate change risks that their constituency is faced with. 
Thus they present empirical evidence to support the \emph{riding the wave} theory~\cite{ansolabehere1994riding}.
The theory suggests that political campaigns are more successful when following the topics that are currently in the public eye.
This implies that the definition of a political agenda is a bottom-up process in which political actors respond to the priorities of potential voters~\cite{Klver2016SettingTA}.
\citet{brooks_ad_framing_2020} study how gender equality, girl power, and climate change are framed in 150 ads. They found that 84.4\% of climate change ads use a `loss frame', highlighting the consequence of inaction rather than the gains of action. 

Understanding climate change related issues on Meta's platforms, however, is a more sparse topic in the literature. 
This is likely related to Meta's restrictive policies on data transparency.
As such, when it comes to Facebook studies have focused on: how climate change denial circulates within public pages~\cite{bloomfield2019circulation}, how NGOs frame the issue~\cite{vu2021social}, or small online experiments to test interventions that address fake news regarding climate change and other topics~\cite{lutzke2019priming}.
Unlike Twitter, Meta has not had an API for researchers to investigate the content on their platform until the release of the Ad Library~\cite{FB-social-issues-classified}.
However, only ads relating to social issues, elections or politic are required to include information about who paid for them.
While the Ad Library has yet to be used to study climate change advertisement it has been explored for various issues and used to studying adjacent policy issues such as COVID-19 and immigration.
\citet{fowler2021political} in particular, analyse the difference in how US politicians advertise on Facebook compared to television and study a range of different political issues. Their findings suggest that politicians are less likely to talk about controversial public policies on Facebook but instead prioritise promotional, valence-oriented ads to activate their existing supporters.
\citet{edelson2019analysis} were some of the first to use the Ad Library and studied how political advertisers in the US disseminate political messages.
They, further, compared the results from the Meta Ad Library to ads disseminated on platforms such as Google and Twitter and pointed out how advertisers can `intentionally or accidentally' bypass the political advertising archive.
Similar studies have also been performed for ads run by political advertisers in Germany~\cite{medina2020exploring}.
Other studies have taken further steps and analyzed the effects of political online advertising by combining Meta's Ad Library data with additional data sources. 
For instance, \citet{jamison2020vaccine} investigated vaccination-related advertising prior to COVID-19 and found that anti-vaccine advertisers were successful in publishing advertisements with low costs but high user impressions.
Similarly, \citet{mejova2020covid} examined narratives around COVID-19 through ads promoted on Facebook, and found several instances of possible disinformation and misinformation, ranging from conspiracy theories regarding bioweapons to unverifiable claims made by politicians.
Lastly, \citet{capozzi2021clandestino} used data from the Ad Library to study immigration stances in Italy, and built a pro- and anti-immigration classifier to dig into which audiences these ads target.

Our work aims to transfer the above established methodologies to analyze ads collected from the Meta Ad Library to the domain of climate change. 
Consequently bridging an important gap in the existing literature, namely studying political communication about climate change on Meta's platforms.

\section{Methodology} 
This section describes how we collect and clean ads from the Meta Ad Library, and how we identify ads relating to climate change topics.

\subsection{Data collection}

Meta's Ad Library API allows us to query data for each politician or political organization by specifying the \emph{page IDs} of the desired Meta pages. 
To identify Congress members we use a publicly available data set~\cite{Congress-legislators} that lists the current members of Congress\footnote{In fact, the data set lists 538 members of Congress and not the precise number of 535 currently serving members. The length is somewhat arbitrary, because members that left Congress may not be updated immediately.} and includes further information such as party affiliation and social media accounts. 
To find the Meta ad account \emph{page IDs} of the politicians we use Meta's own Ad Library report, which lists all political Meta ad accounts since 2018~\cite{Facebook3}.
We retrieve \emph{page IDs} by searching the Ad library using the name of individual politicians.
If a politician has several ad accounts registered\footnote{For instance, Congressman French Hill has pages "Congressman French Hill" and "French Hill for Arkansas".}, we take the page with the highest ad spend.
We use this filtering process to minimize errors and identify the main ad accounts in use by politicians.
We ensure integrity of the pages, by filtering out pages that do not have Meta's official page verification badge \cite{facebook6}\footnote{Through manual revision we filtered out third-party pages such as “Friends of Derek Kilmer" and "Corrupt Thom Tillis".}.
During the manual revision, we removed, replaced, and added several Meta page IDs of politicians.
Overall, we were able to collect the page IDs of \emph{520} Congress members\footnote{A list can be found on \url{https://github.com/lrnz-asnprs/political-ad-api/blob/main/src/data_sets/US_legislators_page_ids_2021.csv}.}.
Fifteen members of Congress did, according to the Ad Library report, not have a Meta ad account linked to their Meta social media accounts.

With this final list of politician page IDs, we crawl the Ad Library API to collect all ads run on these pages from May 2018 until November 2021.
This results in a total of 602,546 ads over all of Meta's platforms.
Even though Meta officially announced that the labeling of political advertisements started in May 2018 and is not backdated~\cite{techcrunch-facebook-api-start}, our data set contains advertisements prior to this date (692 ads in total).
We keep these as in the data set but disregard them when analyzing temporal trends.

\subsection{Filtering for climate-related ads}
In order to determine which advertisements were concerned with the topic of climate change, we adopt a keyword filtering approach that has been used in previous studies~\cite{cody2015climate, yu2021tweeting}. In particular, we filter the collected ads according to a \emph{query} suggested by~\citet{yu2021tweeting}. This query has previously been applied to identify Tweets about climate change.
The query consists of several words related to climate change linked by logical operators: \emph{"climate OR (global AND warming) NOT (business climate OR economic climate OR biz climate OR tax climate OR regulatory climate)"}.
Applying the query to our dataset identifies 19,176 advertisements ($\sim 3\%$) to be related to climate change.
In total, 153 unique Meta page IDs of politicians are included in this subset. Hence, 367 out of the 520 Congress members do not advertise about climate-related topics.

To validate the accuracy of our keyword-filtering approach we compare it to a machine-learning tagging method.
Using an natural language inference-based zero-shot text classifier\footnote{The zero-shot classifier can be found on \url{https://huggingface.co/facebook/bart-large-mnli}.}~\cite{lewis2019bart} we apply it on our total set of ads.
We do so by specifying the label: "climate" and set a probability score threshold of 0.85.
We find that the zero-shot approach classifies 98.4\% of the 19,176 ads identified by the keyword approach to be climate-related. 
The classifier also finds 17 additional climate ads that the keyword approach did not identify, however, after a manual inspection we deem none of them to be climate-related.
As such, we continue using the ads identified by the key-word approach.

\section{Results}

We analyse in total 602,546 ads and focus on understanding: 1) Which politicians drive the topic of climate change and how the temporal dynamics and characteristics of ads run by Democrats and Republicans differ? 2) How the content of ads differs. And, 3) Which factors (demographic, political, and geographical) impact ad performance.

\subsection{Characterizing the total data set} \label{total-data set}

We begin by describing the characteristics of the total dataset before turning to the subset of climate-related ads.
Out of our 520 Congress member large data set, 264 are Democrats ($50.7\%$), 254 are Republicans ($48.8\%$) and two are classified as Independent ($0.5\%$, Senators Bernie Sanders and Angus King).
Out of the total ads, 481,144 ($80\%$) are linked to Democrats (here we count Bernie Sanders and Angus King as Democrats as they both caucus with the Democratic Party) and 121,402 ($20\%$) belong to Republicans.
For each ad, Meta provides the number of impressions (how many times an ad has been seen), and how much funds have been spent on the ad. 
Unfortunately, Meta does not give us the exact number of impressions, or money spent per ad, instead, they provide ranges. 
Impressions are grouped into eight groups ($\le$ 1 K, 1-5 K, 5-10 K, 10-50 K, 50-100 K, 100-500 K, 500 K - 1 M, and  $\ge$ 1 M impressions), while money spent is given in the ranges ($\le$ \$100, \$100-\$499, \$500-\$999, \$1000-\$5000, \$5000-\$10,000, \$10,000-\$50,000, \$50,000-\$100,000, and $\ge$ \$100,000).
To understand the total spend and impressions generated by individual politicians we aggregate their ads by taking the sum of the average endpoints of each individual ad.
For the bottom and upper brackets (e.g. spend brackets $\le$ \$100 and $\ge$ \$100,000) we respectively take the upper or lower limits.
For example, if a politician has one ad in the $\le$ \$100 and three ads in the bracket \$100-\$499 we calculate the total spend as $\$100 + 3 \cdot (\$499 + \$100)/2 =  \$998.5$.

\begin{figure}
    \centering
    \includegraphics[width=8cm]{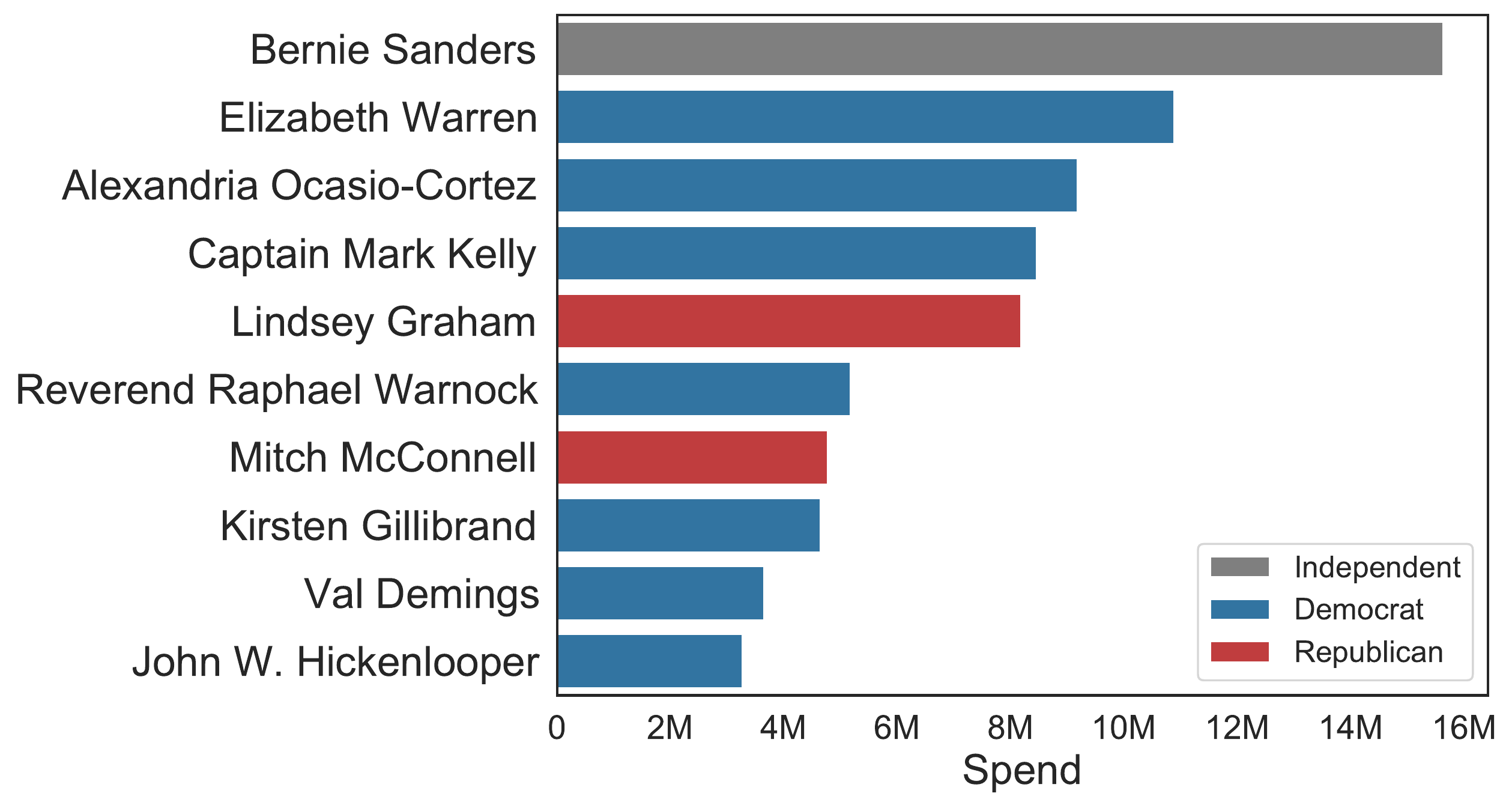} \\
    \includegraphics[width=8cm]{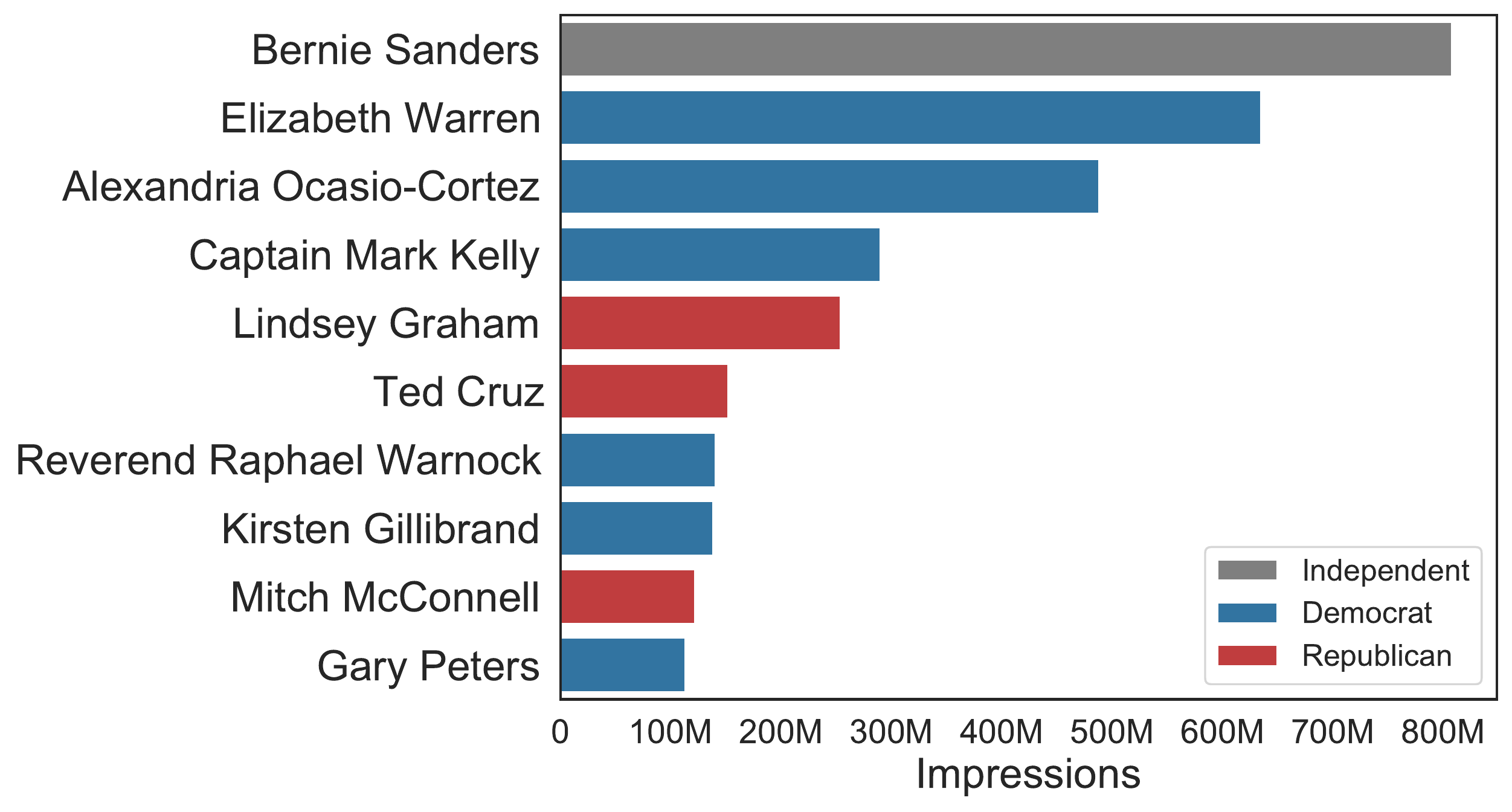}
    \caption{Top 10 advertisers of the total dataset sorted by total spend [\$] (top) and cumulative impressions (bottom). Politicians are colored according to their status in Congress.}
    \label{fig:comparison-top-overall-advertisers}
\end{figure}

Fig.~\ref{fig:comparison-top-overall-advertisers} depicts the overall top 10 advertisers by total spend and impressions.
The top 10 advertisers by spend are dominated by Democrats ($n_{D}$ = 7), with only two Republicans among top advertisers.
Similarity the politicians that received most impressions are also Democrats.
Bernie Sanders is the Congress member which spent most money on ads ($\sim$ \$15M) and generated the most impressions ($\sim$ 800M) during our observation period.
However, comparing spend to impressions we find that there is not necessarily a linear relationship between the two.
For instance, the Republican Ted Cruz appears among the top 10 with the most impressions, although he is not among the top spenders---his ads effectively generate more impressions per dollar.
There can be multiple explanations for this, he can have a bigger follower-base, his ads can be 'better', or his ads can be amplified more strongly by Meta's algorithms.
With the current data from the Ad Library, it is impossible to pinpoint what underlying factors contribute to this difference.

\subsection{Politicians driving the topic of climate change}

Next, we turn to advertisements that are related to climate change.
The subset of climate ads accounts for $\sim3\%$ ($n_{C}$=19,176) of all ads with only 153 (29\%) members of Congress advertising about climate-related issues.
The large majority of climate-related ads generate few impressions and are inexpensive (see Table~\ref{tbl:climate-ads-stats}).
In fact, 96\% of climate ads cost less than \$500 and 67\% of ads generated fewer than 1,000 impressions.
However, there are 8 climate ads with more than 1 million impressions, while no ad costs more than \$100,000.
Comparing these numbers to non-climate related ads, we find that 91.6\% of ads costed less than \$500, while 54\% of ads generated less than 1,000 impressions (a 13\% difference from climate-related ads).
Overall, politicians tend to spend fewer funds on climate-related ads, which subsequently generate fewer impressions.

\begin{table}
\begin{tabular}{l l l}\\
& Climate ads & Non-climate ads \\
& ($n_{C}$=19,176) & ($n=583,368$)\\
\emph{Impressions}\\
\hline 
$\le$ 1 K & 67\% (12,909) & 54\% (314,993) \\ 
1-5 K & 19\% (3681) & 22\% (129,169) \\ 
5-10 K & 5\% (953) & 7\% (43,435) \\ 
10-50 K & 7\% (1251) & 11\% (66,434) \\ 
50-100 K & 1\% (199) & 3\% (14,638) \\ 
100-500 K & 0,8\% (160) & 2\% (12,849) \\ 
500 K - 1 M & 0,08\% (15) & 0.2\% (1258) \\ 
$\ge$ 1 M & 0,04\% (8) & 0.1\% (592) \\
\emph{Spend}\\
\hline 
$\le$ \$100 & 86\% (16,455) & 76\% (443,869) \\ 
\$100-\$499 & 10\% (1940) & 15\% (90,290) \\
\$500-\$999 & 2\% (372) & 4\% (21975) \\
\$1000-\$5000 & 2\% (339) & 4\% (22,697) \\
\$5000-\$10,000 & 0,2\% (35) & 0.5\% (2817) \\
\$10,000-\$50,000 & 0,2\% (34) & 0.3\% (1615) \\
\$50,000-\$100,000 & 0,005\% (1) & 0.02\% (94) \\
$\ge$ \$100,000 & 0\% (0) & 0.004\% (21) \\
\hline
\end{tabular}
\caption{Distribution of climate ads and all ads in ranges of impressions (top) and spend (bottom). Numbers are summarized over individual ads.} \label{tbl:climate-ads-stats}
\end{table}

\begin{figure}
    \centering
    \includegraphics[width=8cm]{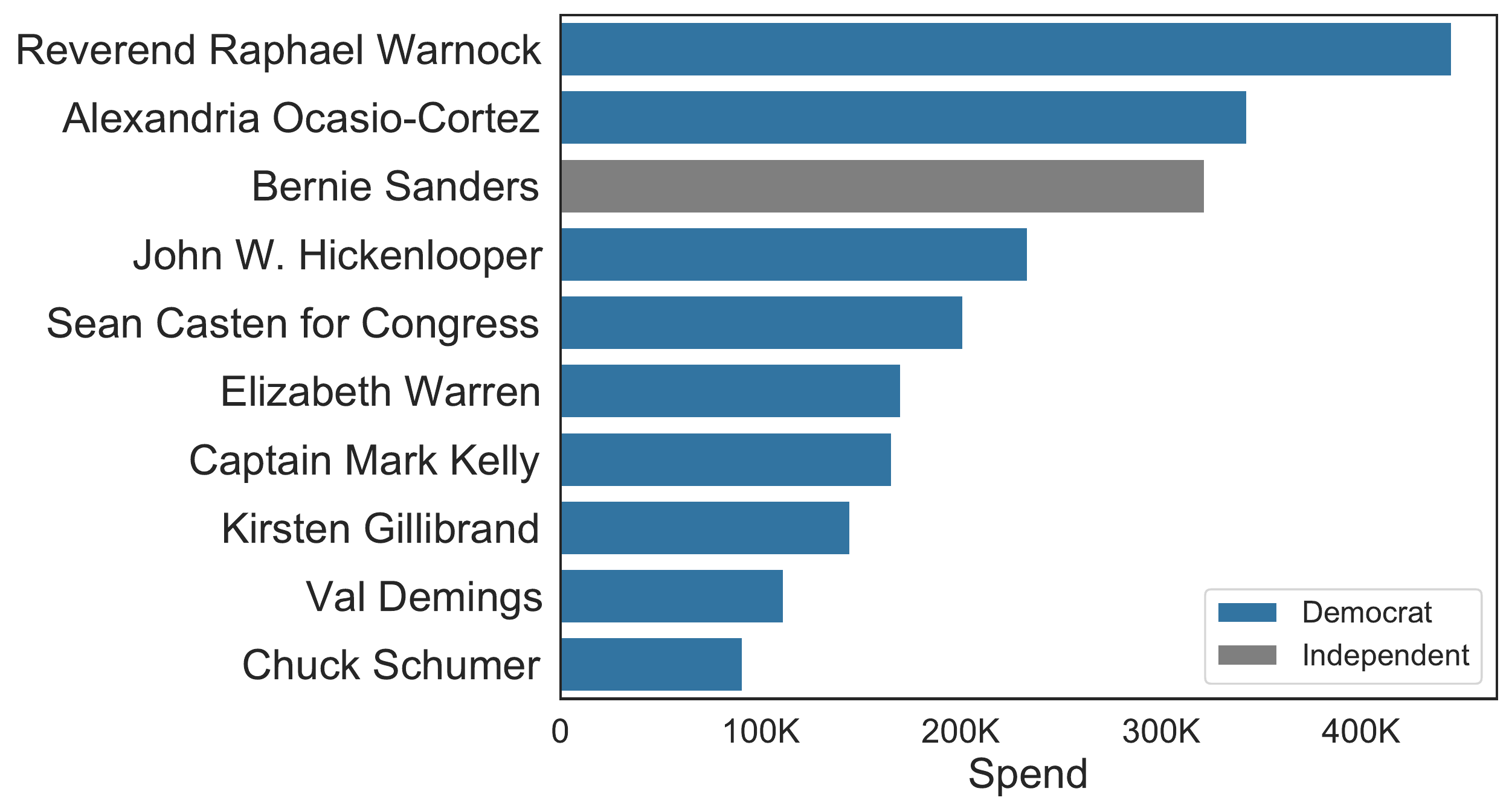}\\
    \includegraphics[width=8cm]{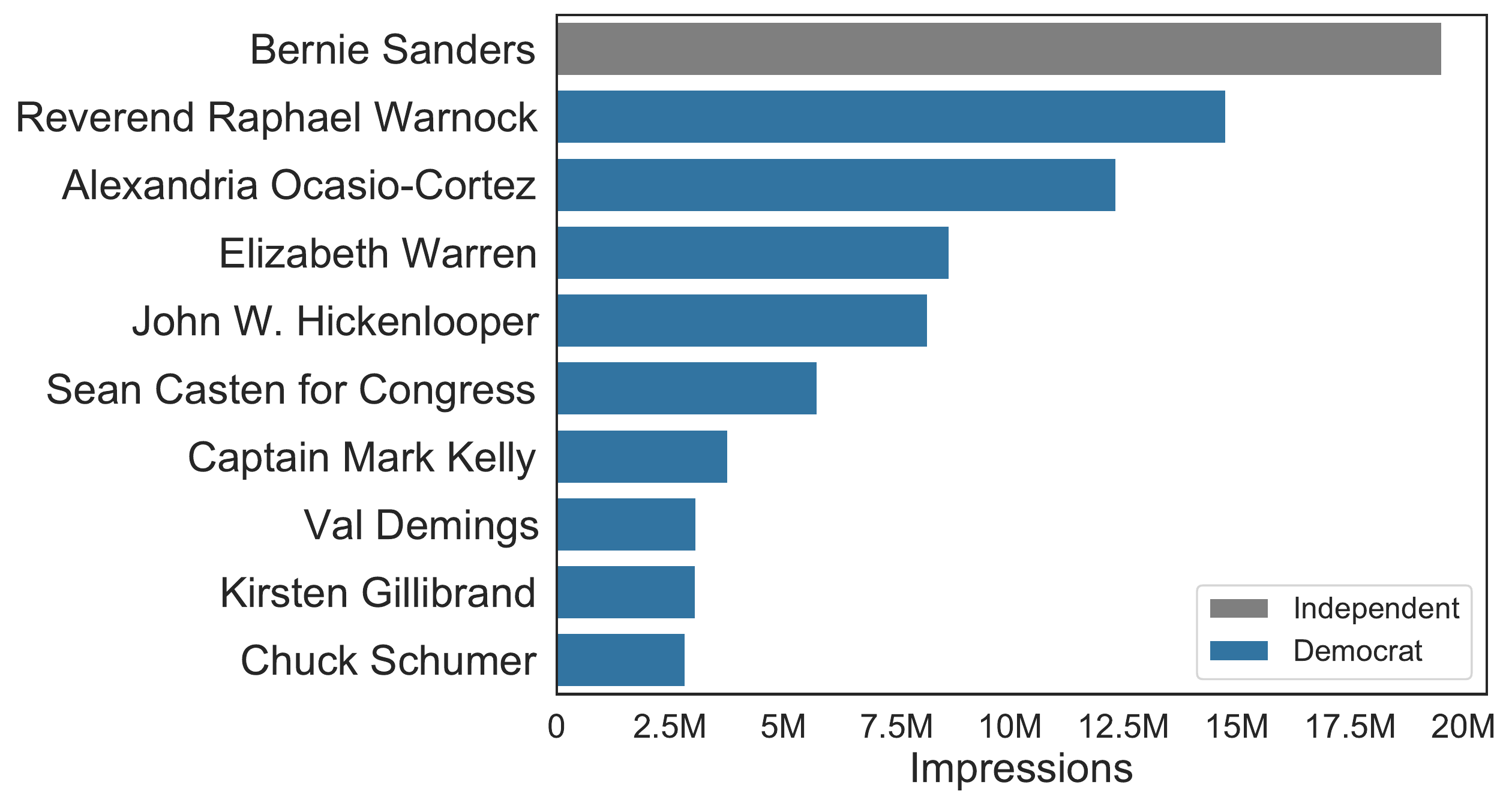}
    \caption{Top 10 climate advertisers sorted by total spend [\$] (top) and the cumulative number of impressions (bottom).}
    \label{fig:comparison-climate-advertisers}
\end{figure}

Out of the 153 Congress members that run climate-change related ads, 140 are Democrats ($92\%$), while only 13 are Republicans ($8\%$).
Fig.~\ref{fig:comparison-climate-advertisers} shows that the top 10 climate advertisers differ from the top 10 overall advertisers.
Most notably, there are no Republicans among the top climate advertisers.
All of the top 10 politicians according to spend also appear in the top 10 according to impressions, however in a different order.
While Bernie Sanders, for instance, ranks third by spend he is the politician with the highest number of impressions for climate-related ads.
Further, when we compare the impressions generated by the top 10 climate advertisers we find that they account for 72\% of the total number of impressions for climate ads.

Looking at what proportion climate ads account for out of the total number of ads for each politician we find that only $15\%$ of politicians that talk about climate change have a share of climate ads higher than $10\%$.
The three politicians (with at least 100 ads) with the highest fraction of climate change related ads are Jimmy Panetta (40\%), Ed Perlmutter (37\%), and Jared Huffman (36\%).
This reveals that only a small share of politicians emphasize the topic, and none run exclusively on a climate platform. 
Overall, few politicians run climate-related ads, and out of those, climate ads only account for a small fraction of their total ads.

\subsection{Temporal dynamics of climate ads}
\begin{figure*}
    \centering
    \includegraphics[width=0.49\textwidth]{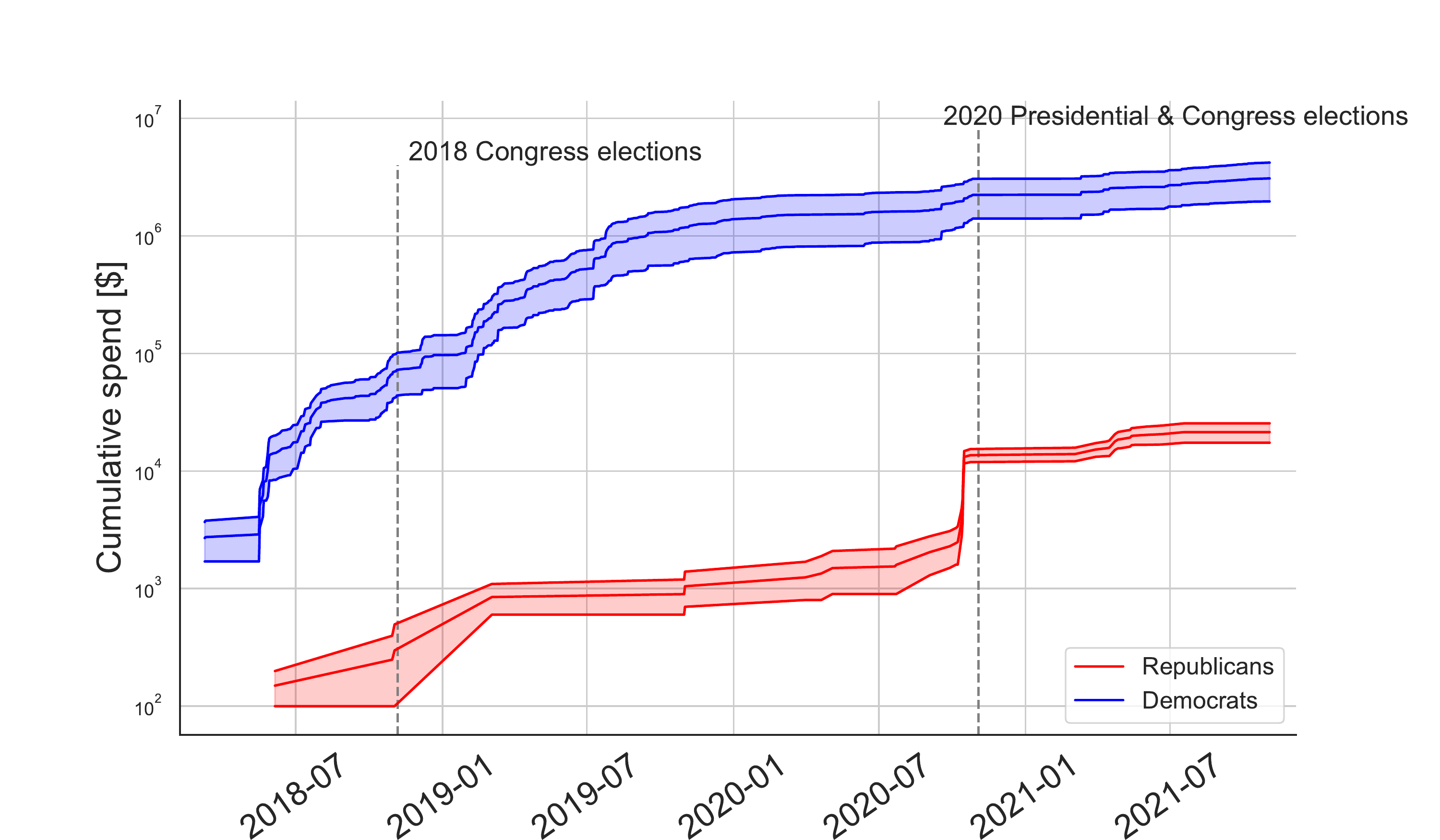}
    \includegraphics[width=0.49\textwidth]{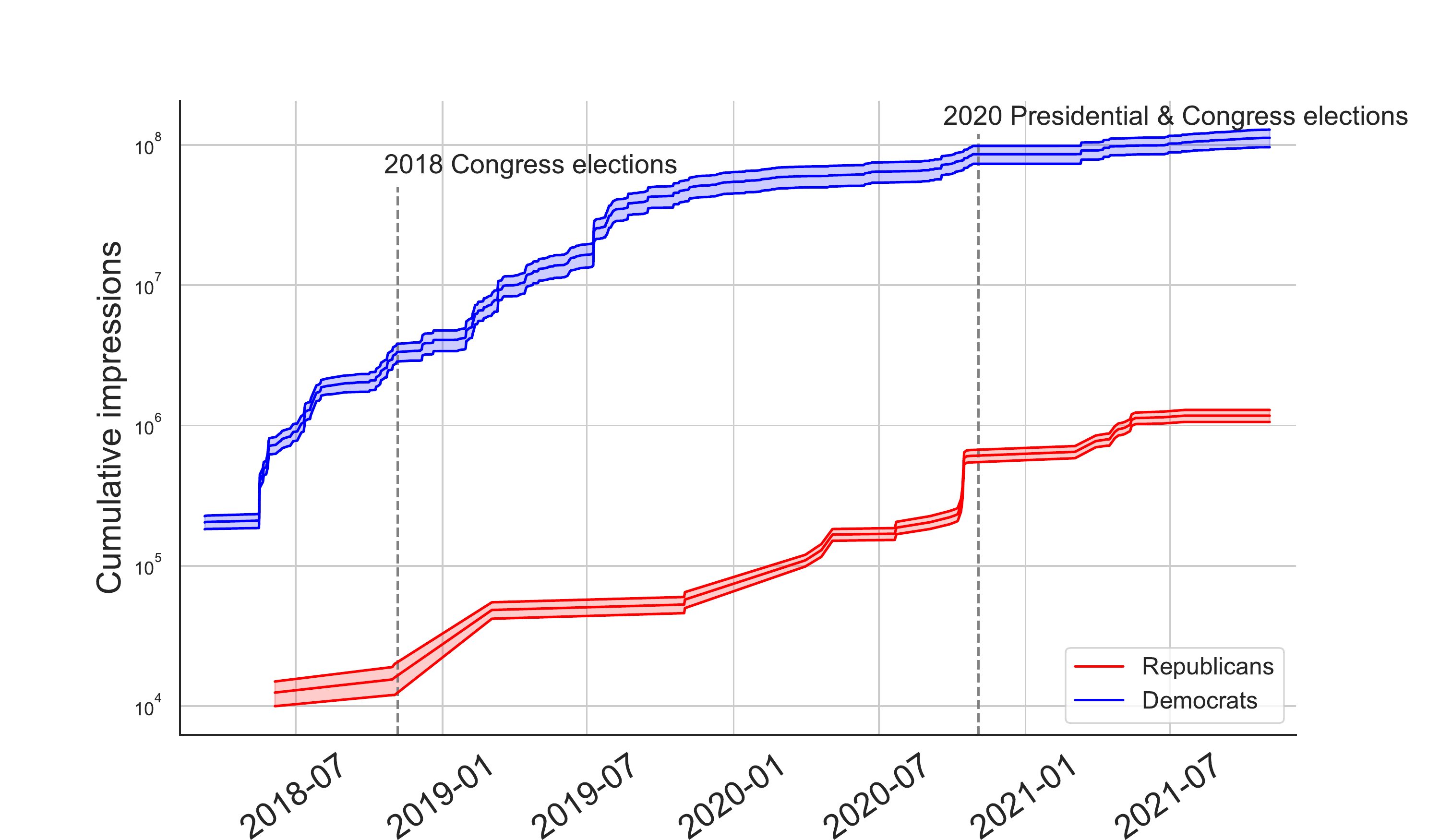}
    \caption{Temporal dynamics of climate-related ads. (Left) Cumulative spend [\$] by political parties since the launch of Meta's Ad Library in 2018. Here we have assigned Bernie Sanders and Angus King as Democrats. Note, that we show the lower, average, and upper bounds of the spend range. The upper and lower bounds are estimated by summing up all upper- and lower-range values. Further, please note that due to the large differences in between parties we use logarithmic scaling on the y-axis. (Right) Cumulative impressions, where logarithmic scaling is used to compare the political parties.}
    \label{fig:time-climate-ads}
\end{figure*}

To further quantify the ecosystem around climate-related ads we focus on understanding when individual ads are run, how many impressions they generate, and how funds are spent over time.
First, we split climate-related ads according to political parties and find that Democrats account for 99.6\% (19,107) of all climate-related ads, with Republicans having, in total, only run 69 ads (0.04\%) in the period from March 2018 to November 2021.
Fig.~\ref{fig:time-climate-ads} (left panel) shows the temporal dynamics of climate-related ads and illustrates significant differences between Democrats and Republicans.
While Democrats have cumulatively spent more than \$3 million since mid-2018, Republicans have only spent \$21,000 on climate-related ads---the difference is more than two orders of magnitude.
The difference in impressions shows a similar picture (Fig.~\ref{fig:time-climate-ads}, right panel).
Democrats have generated a cumulative amount of 112 million impressions, while Republicans have generated a total of 1,17 million impressions.
Similar findings were found for Twitter, where Democrats have been found to tweet more frequently about climate change related topics~\cite{yu2021tweeting}.

Looking at the temporal dynamics, we observe a sudden increase in Republican spend and impressions in the months prior to the 2020 presidential and congressional elections.
Overall, Republican candidates spent more than half of their total cumulative funds within a two-month period.
However, the jump in spend is exaggerated by the logarithmic scale of the graph.
While Democratic candidates also spend more prior to the 2020 elections, their increase is less visible. Democrats spent 18\% of their total funds during the same period prior to the 2020 elections. For Democrats, the relative share of climate ads impressions has grown over time. Pre-2021, impressions stemming from climate ads constituted on average less than 3\% of all impressions per month, while from 2021 and onward, they generated, on average, 6\% of monthly impressions. Interestingly, there is relatively higher spend and impressions on climate ads after election dates.

Spend and impressions alone do not give us a complete image of the dynamics, as there are large differences in how many ads candidates from the two parties run ($n_{D}=19,107$, $n_{R}=69$).
Instead, to quantify differences in how the two parties advertise and how their ads perform, we look at the average spend per ad and the average number of impressions generated per dollar.
Overall, Democrats spend on average \$161 per climate ad, whereas Republicans spend \$310, although there is no significant difference between the two means (two-sample t-test, $t = -1.23$, $p = 0.22$).
However, we find a significant difference in the average impressions per dollar between the two parties (two-sample t-test, $t = -10.98$, $p < 0.001$).
Republicans generate on average 72 impressions per dollar compared to 25 for the Democrats.
In other words, although Republicans advertise much less about climate-related topics they are more successful at generating impressions---they outperform Democrats in generating impressions per dollar by 188\%.

\subsection{Sentiment of ads}
To understand if differences in impressions are caused by how politicians talk about climate change we analyze the content of their ads.
Here, we focus the sentiment of ads, which we calculate using the VADER library~\cite{hutto2014vader}.
Ads which have no associated text (e.g. they only contain a video or image) are disregarded (less than 1\% of ads).
Ads can consist of multiple sentences, to quantify sentiment we score each individual sentence and take the average compound score.
As such, an ad is only negative (or positive) if it consistently uses negative (or positive) language. 

\begin{figure*}[htbp]
    \centering
    \includegraphics[width=\linewidth]{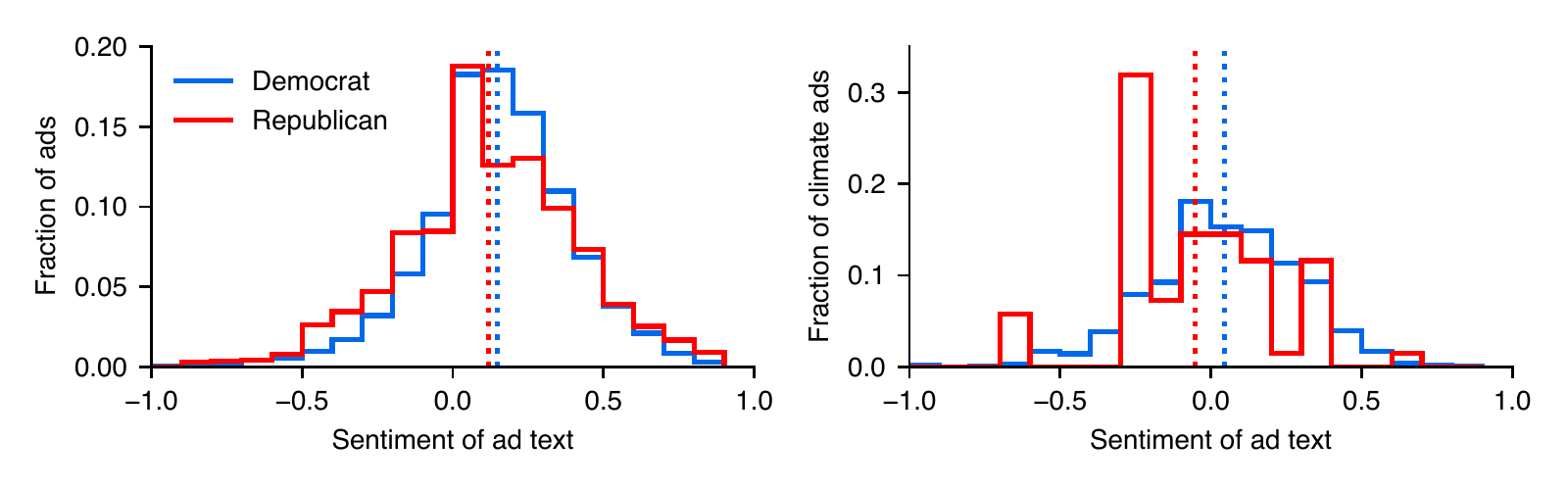}
    \caption{Sentiment distributions. Left, Distributions for all political ads. Dotted lines show the average sentiment. Right, sentiment distributions for ads related to climate change topics. The Republican distribution is more jagged due to a lower number of climate related ads.}
    \label{fig:sentiment}
\end{figure*}

Fig.~\ref{fig:sentiment}, left panel, shows the sentiment distributions for all ads.
While the average values are very similar and positive for both parties, 0.15 for Democrats, and 0.12 for Republicans, the distributions are different. 
The sentiment distribution for Republican ads is more broad, indicating that their ads are more likely to be ‘extreme’, either being overly negative, or overly positive (tails of the distribution).
Further, calculating the average sentiment for ads in each of the 8 impression groups (see Table~\ref{tbl:climate-ads-stats}) we find that Republicans ads are, on average, always less positive than Democrat ads.

For climate ads (Fig.~\ref{fig:sentiment}, right panel), we see that both parties use nuanced sentiment to talk about climate topics.
However, Republican ads are, overall, shifted towards more negative values and have, on average, negative sentiment values (-0.05).
Democrat climate ads are slightly positive with a mean value of 0.04.
As such, lower, or more negative, sentiment values can be one explanation to why Republican ads generate more impressions per 
dollar. This is in line with recent studies which have shown that negative comments generate more engagement on Facebook~\cite{metz2021likes}, and that algorithmic amplification of this content can lead to great reach~\cite{reuning2022facebook}.

\begin{figure*}
    \centering
    \includegraphics[width=8cm]{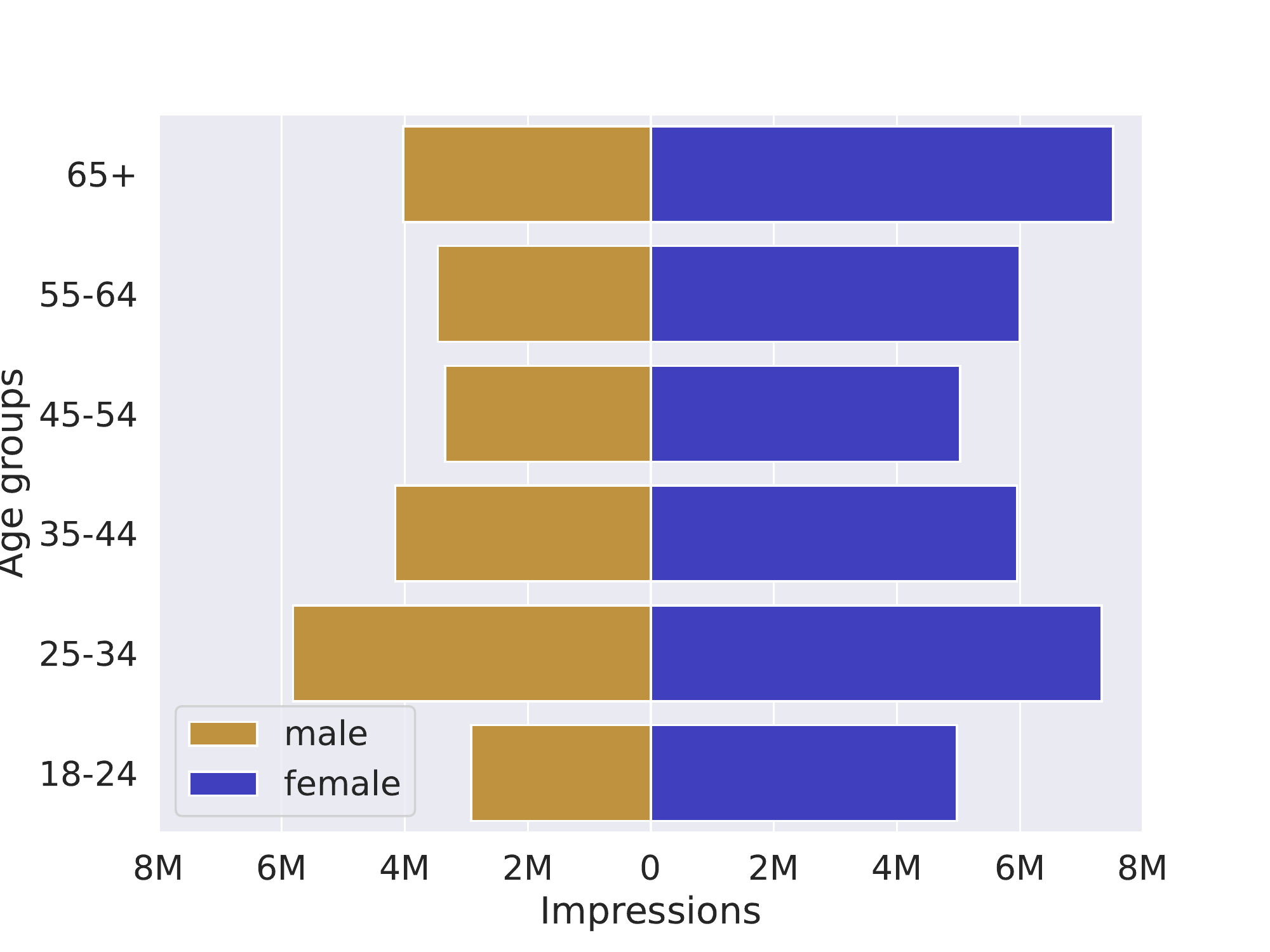} 
    \includegraphics[width=8cm]{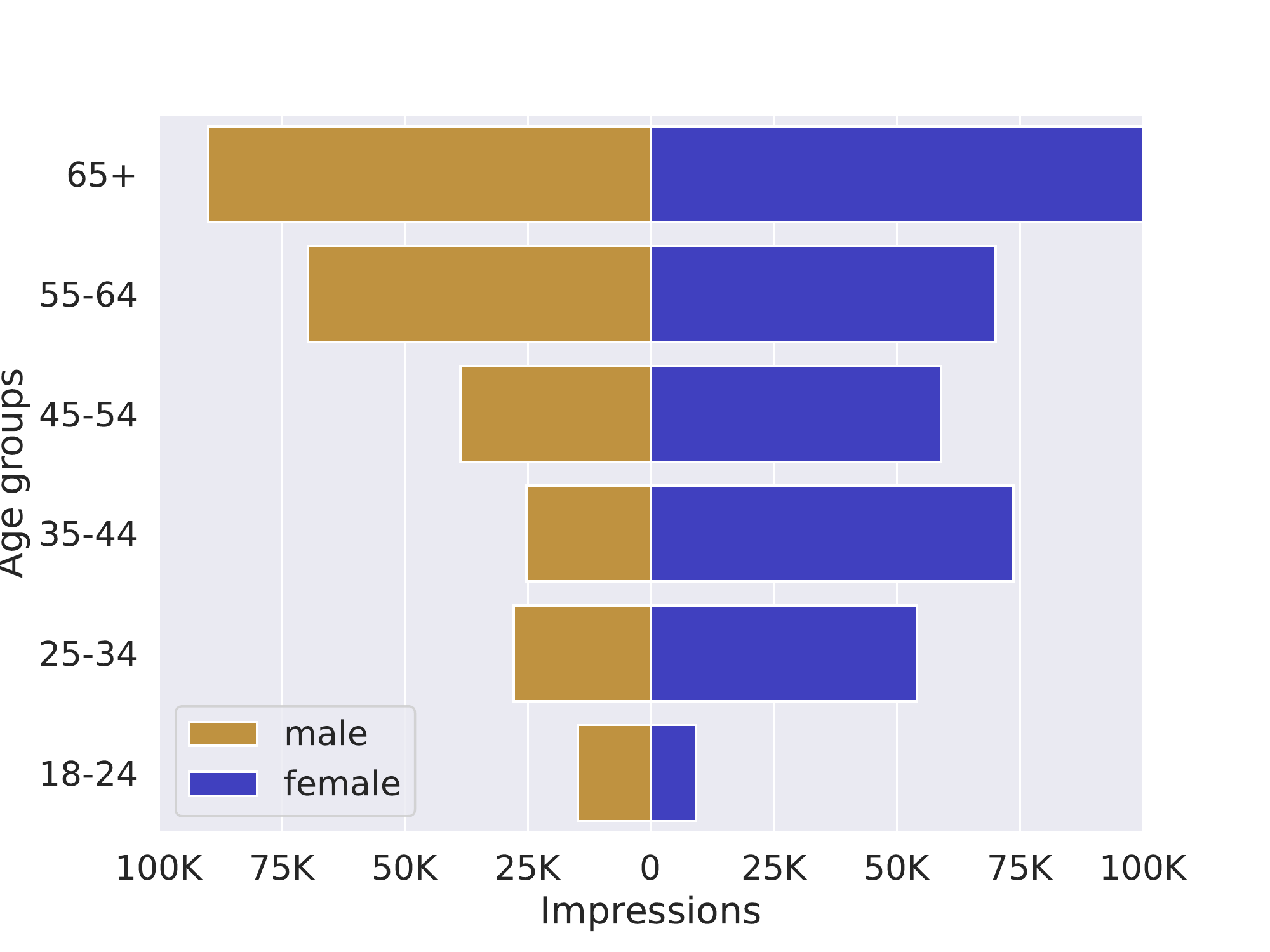}
    \caption{Audience age and gender distributions of climate-related ads run by Democrats (left) and Republicans (right), measured by the number of impressions. The distributions for non-climate ads are similar.}
    \label{fig:climate_ads_demographic_pyramid}
\end{figure*}

\subsection{Audience demographics}
Spend and sentiment might not be the only factors that determines the `performance' of an ad.
Targeting demographics, i.e. which audiences the politicians target their ads towards, might also explain some of the differences.
In the following, we focus on the set of ads for which there is demographic and geographic information for---this covers 13,466 ($70\%$) climate-related ads and 463,402 ads in total ($77\%$). 
Meta's Ad Library contains demographic information about the age, gender (binary), and geographical location (US states) of people who viewed each ad. 
Fig.~\ref{fig:climate_ads_demographic_pyramid} shows what age and gender segments viewed climate-related ads.
Ads run by Democrats tend to be viewed by younger audiences in comparison to their counterparts.
In particular, 25-34-year-olds, whereas the Republicans have an older audience (i.e. older than 55).
The demographic data for Democrats reflect an hour-glass shaped distribution, with a higher fraction of users in the segments 25-34 and +65, while for the Republicans the distribution takes the shape of a reverse pyramid.
However, similar patterns are present in the demographic distributions for all ads.
As such, climate-related ads do not deviate from the demographic distributions of all other ads run by politicians from the two parties.
A general trait for demographic distributions is that there are more female impressions. 
This can be explained by the overall distribution of Meta's socials being skewed towards female demographics~\cite{statista-facebook-demographics, statista-instagram-demographics_US}.
Further, it is interesting that both parties have a high fraction of their impressions in the 65+ year segment, despite the fact that this segment only accounts for about 5\% of all Facebook users and 2.1\% of Instagram users \cite{statista-facebook-demographics, statista-instagram-demographics}.

To understand the effects of demographic factors in more detail we use the information to build a model.
Here, we only focus on ads for which the lower bounds of impressions and spend are non-zero, and ads for which we have demographic information for.
We do this because we are interested in knowing which factors cause an ad to be successful, not which factors cause an ad to be unsuccessful.
In total, this leaves us with a dataset of approximately 25\% of all ads.
Further, to account for heterogeneity in ads we build two models, one for each political party.
As model type we focus on linear models (LASSO) due to their interpretability. 

\begin{figure}
    \centering
    \includegraphics[width=\linewidth]{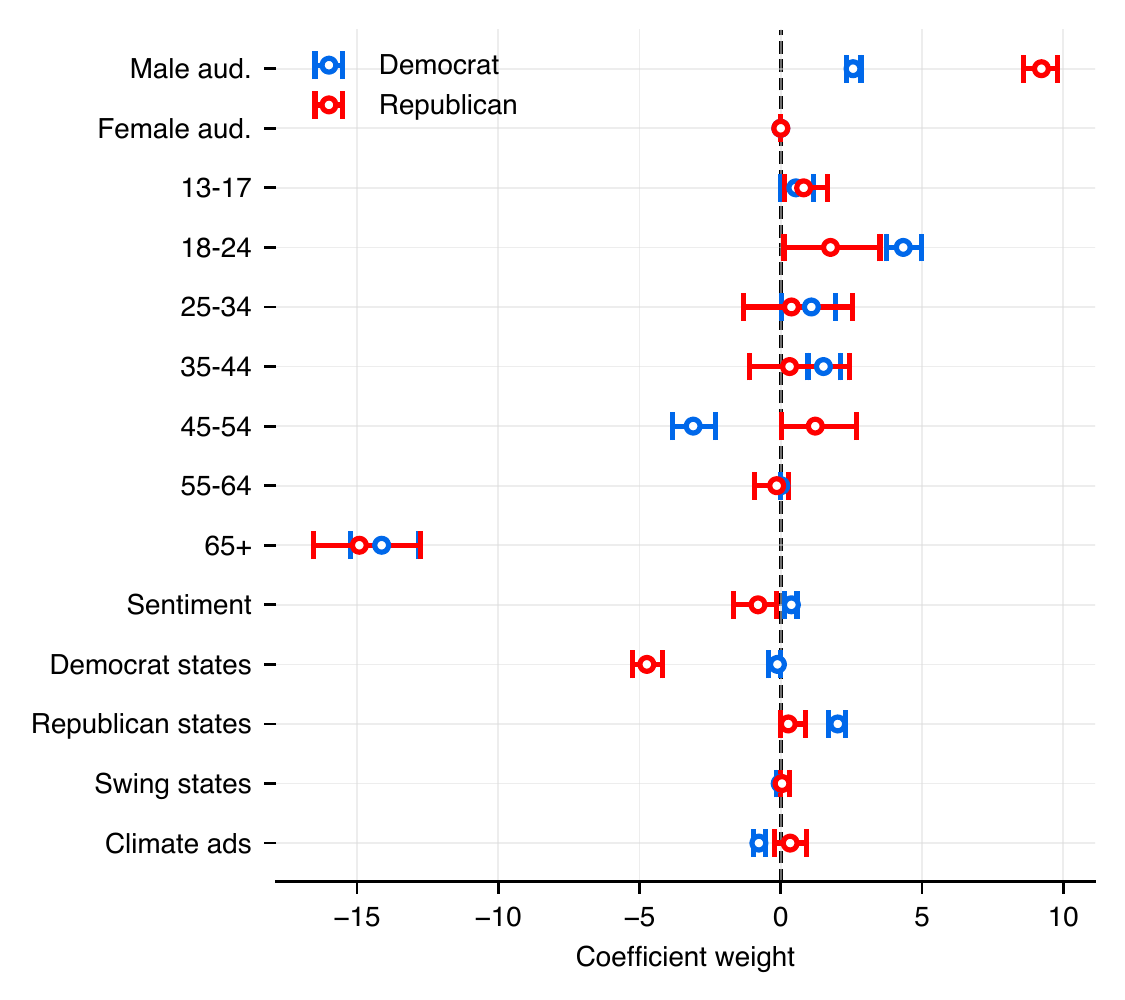}
    \caption{Linear models for understanding factors behind impressions per dollar. Figure shows the coefficient weights for individual party models covering all ads. Error-bars show 95\% confidence intervals, estimated from 100 bootstrapped samples.}
    \label{fig:models}
\end{figure}

Fig.~\ref{fig:models} shows coefficient weighs for the models.
We find that the models do not have strong predictive power (correlation coefficient r = 0.476 for the Democrat model and r = 0.297 for the Republican model).
However, the goal is not to get a perfect prediction, rather we want to understand the factors that drive impressions per dollar.
We find that impressions for Republican ads are mainly driven by male audiences, while for Democrats it is 18-24 year olds, followed by male audiences.
(Female audiences have zero weight in the model as this variable is redundant. Female and male audiences sum to one so the `female audience' variable does, model-wise, not contain new information.)
Interestingly, the factors limiting impressions per dollar are for both models 65+ year old demographics.
Geographic regions only play a minor role, but Republican ads perform worse in Democrat states, while Democrat ads have a positive contribution from Republican states. 
(We have chosen to aggregate US states into three categories, Democrat, Republican and swing states. We define swing states as a state where either party got less than a 5\% win margin in the 2020 presidential election. Swing states are: Arizona, Florida, Georgia, Michigan, Nevada, North Carolina, Pennsylvania, and Wisconsin.)
When it comes to ad sentiment the models show that positive sentiment has only a minor positive contribution for Democrat ads, while the weight is negative for Republican ads. 
This means, that Republican ads with positive sentiment score perform worse; the more negative an ad is the better it performs.
Lastly, climate related ads also have a minor contribution to ad performance.

\begin{figure}
    \centering
    \includegraphics[width=\linewidth]{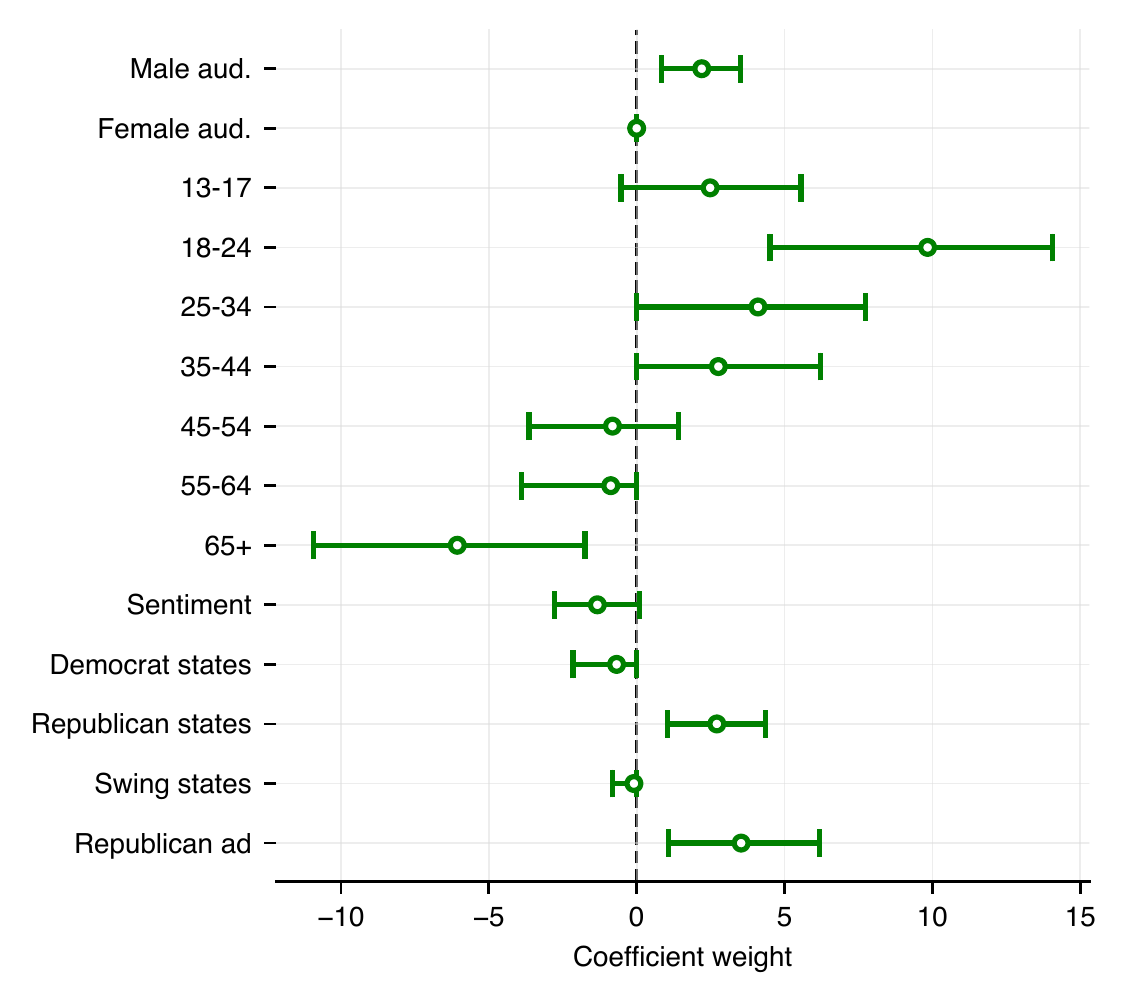} 
    \caption{Coefficient weights for linear models that explains impressions per dollar for climate ads. Error-bars show 95\% confidence intervals, estimated from 100 bootstrapped samples.}
    \label{fig:model_climate}
\end{figure}

Similar to the models in Fig.~\ref{fig:models}, we look at which factors explain impressions per dollar for climate ads (Fig.~\ref{fig:model_climate}).
Here, we only build one model because there are so few Republican climate ads.
This model has a better fit (correlation coefficient r = 0.521), and reveals that the factors which drive impressions per dollar for climate ads are predominantly 18-24 and male audiences. 
The main limiting factor is again 65+ year old audiences.
However, the model also reveals that climate ads viewed in Republican states and ads run by Republicans get more impressions per dollar.

\section{Discussion}

To our knowledge, our work presents the first quantitative study of how US politicians advertise climate change-related topics.
We focus on ads run on Meta's (formerly Facebook) platforms between the period from May 2018 to November 2021 and find almost 20,000 climate-related ads, indicating that climate-change topics only play a minor role in political advertisement ($\sim3\%$ of all ads).

Our work is limited by multiple factors, including data issues, and algorithmic confounders.
First, both impressions and spend are reported as range values instead of being specified as a precise number, which introduces a factor of inaccuracy. 
We have tried to account for it by calculating averages, and showing the range of possible spend (e.g. Fig.~\ref{fig:time-climate-ads}).

A second shortcoming is that Ad Library does not contain any information about the intended targeting of the advertisers, it only shows which demographics ultimately saw the ads.
These two are not the same, especially given the myriad of ways advertisers can target their audiences on Meta, and how algorithms on Meta decide who to show ads to~\cite{ali2019discrimination}.
Having this information at hand would allow us to investigate \emph{to whom} politicians want to speak about climate change and eventually allow us to draw more relevant conclusions about the rationale behind their advertising strategies.
Additionally, it could help us quantify the skew `algorithmic targeting'~\cite{thorson2021algorithmic} introduces in the data.
Meta has previously released such data~\cite{Facebook5}.
In the wake of the 2020 presidential elections, they released a dataset that included targeting information of all political ads that were run in the 90-day period preceding the elections on November 3rd, 2020.
As such, it raises the question of why Meta is not revealing targeting specifications for all political advertisements.
Even though such data might reveal campaign strategies of various political actors, the gain in transparency for research and policy-making would be immense. 

\begin{figure}[!htbp]
    \centering
    \includegraphics[width=\linewidth]{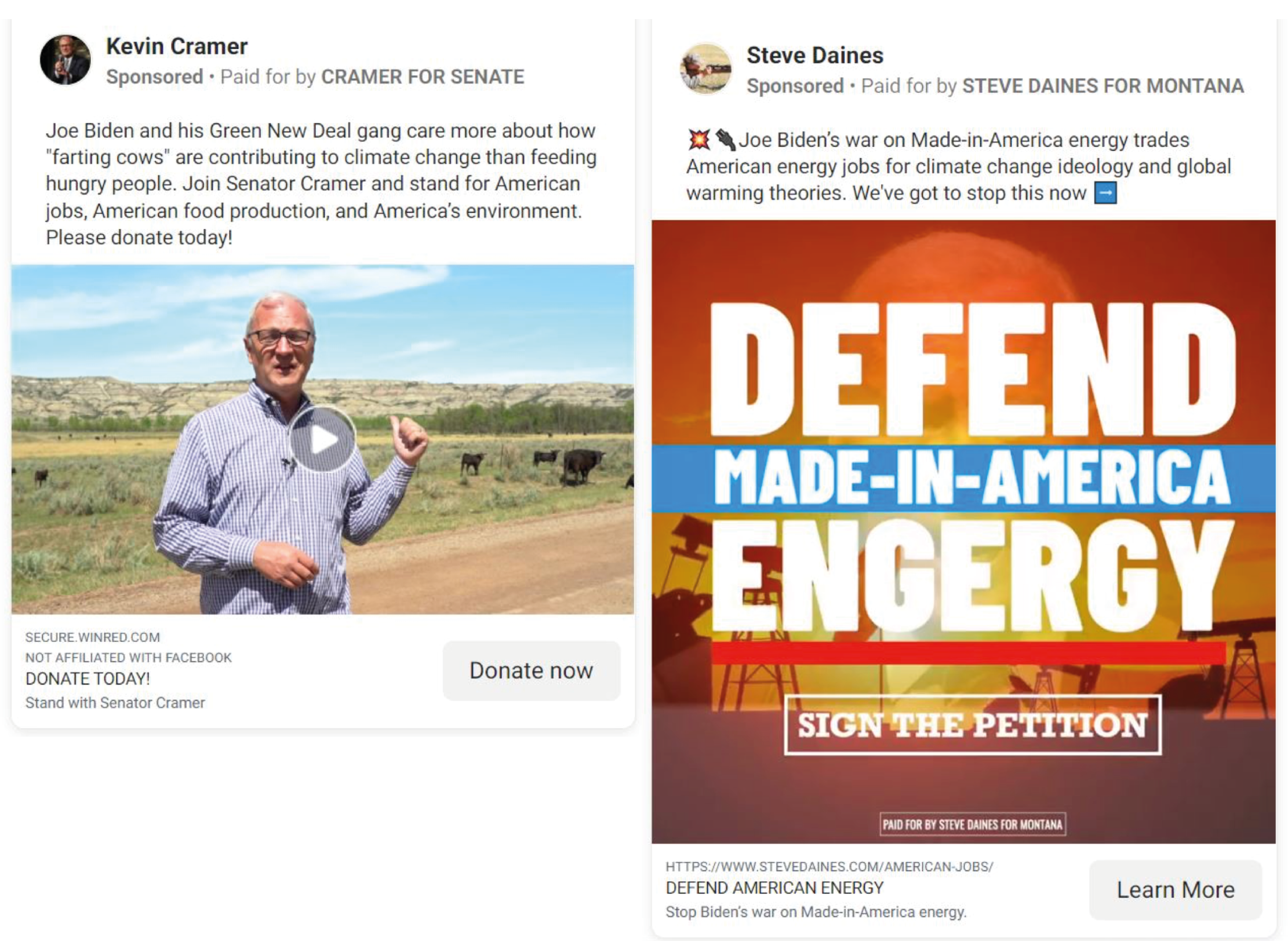}
    \caption{Selected examples of Republican ads marked as climate-related in our dataset, which deny or are critical of human-induced climate change and global warming.}
    \label{fig:ad-examples}
\end{figure}

One major limitation is that the keyword filtering method is limited to textual data, and while a majority of Meta ads contain text, some only contain multimedia content such as videos or images.
Future studies, could include these ads by transcribing the content using speech-to-text or image-to-text methods. 
Further, the keyword filtering method also identifies ads that deny, or are critical of climate change-research as climate-related ads (Fig.~\ref{fig:ad-examples}).
By manually going through a subsample of the ads we found several cases of Republicans being against climate measures and some even denying climate change, calling global warming a `theory'.
We have chosen to keep these in the data as they relate to climate-related topics, just from an opposite perspective.

Lastly, the only metric we have access to in the Ad Library which quantifies the `performance' or `effectiveness' of an ad is the number of \emph{impressions}.
However, that is often not the best way of estimating ad performance, other metrics such as engagements (e.g. number of likes), click-through rates, and relevance scores (calculated by Meta based on the positive and negative feedback they expect an ad to receive from its target audience) would make it easier to understand the impact of climate-related ads.
All these metrics are otherwise available to advertisers through Meta's Ad Manager \citep{Facebook5}.

Despite these limitations, our work reveals interesting insights about the climate-related focus of Congress members.
Our findings show that Democrats noticeably dominate online advertisements about climate change on Meta's platforms.
In fact, there is a substantial difference, as almost no Republicans advertise about climate change ($<0.4\%$ of all climate ads in our data).  
While the imbalance between the parties makes it difficult to draw general conclusions about advertising strategies, it indicates the perceived relevance each party and its potential voters ascribe to the topic of climate change. Unlike TV ads, where the audiences are more politically diverse, Meta ads are targeted and can, therefore, be used by Congress members to convince individuals who share an ideological affinity \citep{fowler2021political}.

The top 10 politicians that talk about climate change are all Democrats, and their ads make up for 72\% of all climate-related ads.
This means that even within the Democratic party, only a few actors focus on climate-related topics.

Comparing the spend and impressions generated by climate ads to all ads, we find that a majority of climate ads fall within the categories of low spend ($<$\$100) and low impressions ($<$1 K).
However, we find that Republican ads not only generate higher average impressions per dollar than Democrats when looking at all ads but also outperform their counterpart when specifically looking at climate-related ads.
This could stem from the fact that Republican ads in general are less positive than Democrat ads.
Manually combing through Republican ads, we also found several cases of politicians being against climate policies and some even denying climate change (Fig.~\ref{fig:ad-examples}).
These examples indicate a political divide of opinions between Democrats and Republicans.
Advertisements that contain polarizing, extreme, and divisive content has been found to generate more attention on Meta's platform~\citep{bakir2018fake}, and on Twitter, where content by the mainstream political right has been found to achieve higher algorithmic amplification than the mainstream political left~\cite{huszar2022algorithmic}.

Looking at audience demographics, we find that Republicans have more impressions in the older segments, while Democrats have a higher share of impressions in younger segments (Fig.~\ref{fig:climate_ads_demographic_pyramid}).
However, our models reveal that these demographics do not necessary drive the number of impressions ads generate per dollar. 
In fact, we find that Republican ad impressions are mainly driven by male audiences, while Democrat ads are predominantly driven by younger audiences (18-24 year olds). 
For both parties older audiences (65+ year olds) actually reduce the number of impressions gained per dollar.
Impressions for climate ads are similarity driven by male audiences, but we also find that audiences from Republican states drive impressions, and that Republican sponsored ads perform better.
However, the models are not perfect, their predictive power is limited by their simplicity, and by the granularity of available data.
Gaining a deeper understanding of ad impressions with respect to social and economic inequalities will require different approaches.

Our work presents new insights and limitations of using Meta's Ad Library for studying how politicians talk about climate change.
While we mainly focused on the US ecosystem around climate ads, one possible avenue of future work could be deeper analysis of the content of ads.
Natural Language Processing techniques can be used to analyze which narratives are used, and whether cases of misinformation about climate change are present in the political advertisement.
The second avenue of future research could be to increase the scope of who is included in the dataset.
While our focus is on the Congress, it would be interesting to include a broader set of politicians (e.g. local politicians), and include NGOs and other active voices in the climate debate in further research. 
Extending the scope to other countries could also be beneficial.
Lastly, comparing how politicians advertise across Meta's different platforms (Whatsapp, Instagram, Facebook) could bring interesting insights. 

\section{Ethical Statement}
The data in this paper is derived from the Meta Ad Library. It contains publicly accessible ads run on Meta platforms by US politicians. Working with social media data carries risks of privacy issues and the right to be forgotten. However, our data analysis is limited to aggregated data presentations and only concerns ads published by public figures.

\small
\bibliography{bibliography.bib}

\begin{thebibliography}{51}
\providecommand{\natexlab}[1]{#1}

\bibitem[{Ali et~al.(2019)Ali, Sapiezynski, Bogen, Korolova, Mislove, and
  Rieke}]{ali2019discrimination}
Ali, M.; Sapiezynski, P.; Bogen, M.; Korolova, A.; Mislove, A.; and Rieke, A.
  2019.
\newblock Discrimination through optimization: How Facebook's Ad delivery can
  lead to biased outcomes.
\newblock \emph{Proceedings of the ACM on human-computer interaction}, 3(CSCW):
  1--30.

\bibitem[{Allcott and Gentzkow(2017)}]{allcott2017social}
Allcott, H.; and Gentzkow, M. 2017.
\newblock Social media and fake news in the 2016 election.
\newblock \emph{Journal of economic perspectives}, 31(2): 211--36.

\bibitem[{Ansolabehere and Iyengar(1994)}]{ansolabehere1994riding}
Ansolabehere, S.; and Iyengar, S. 1994.
\newblock Riding the wave and claiming ownership over issues: The joint effects
  of advertising and news coverage in campaigns.
\newblock \emph{Public Opinion Quarterly}, 58(3): 335--357.

\bibitem[{Auter and Fine(2018)}]{auter2018social}
Auter, Z.~J.; and Fine, J.~A. 2018.
\newblock Social media campaigning: Mobilization and fundraising on Facebook.
\newblock \emph{Social Science Quarterly}, 99(1): 185--200.

\bibitem[{Bakir and McStay(2018)}]{bakir2018fake}
Bakir, V.; and McStay, A. 2018.
\newblock Fake news and the economy of emotions: Problems, causes, solutions.
\newblock \emph{Digital journalism}, 6(2): 154--175.

\bibitem[{Bloomfield and Tillery(2019)}]{bloomfield2019circulation}
Bloomfield, E.~F.; and Tillery, D. 2019.
\newblock The circulation of climate change denial online: Rhetorical and
  networking strategies on Facebook.
\newblock \emph{Environmental Communication}, 13(1): 23--34.

\bibitem[{Bode(2016)}]{bode2016political}
Bode, L. 2016.
\newblock Political news in the news feed: Learning politics from social media.
\newblock \emph{Mass communication and society}, 19(1): 24--48.

\bibitem[{Brooks, Craig, and Bichard(2020)}]{brooks_ad_framing_2020}
Brooks, M.~E.; Craig, C.~M.; and Bichard, S. 2020.
\newblock Exploring Ads of the World: How Social Issues Are Framed in Global
  Advertisements.
\newblock \emph{Howard Journal of Communications}, 31(2): 150--170.

\bibitem[{Capozzi et~al.(2021)Capozzi, De~Francisci~Morales, Mejova, Monti,
  Panisson, and Paolotti}]{capozzi2021clandestino}
Capozzi, A.; De~Francisci~Morales, G.; Mejova, Y.; Monti, C.; Panisson, A.; and
  Paolotti, D. 2021.
\newblock {Clandestino or Rifugiato? Anti-immigration Facebook Ad Targeting in
  Italy}.
\newblock In \emph{Proceedings of the 2021 CHI Conference on Human Factors in
  Computing Systems}, 1--15.

\bibitem[{Cody et~al.(2015)Cody, Reagan, Mitchell, Dodds, and
  Danforth}]{cody2015climate}
Cody, E.~M.; Reagan, A.~J.; Mitchell, L.; Dodds, P.~S.; and Danforth, C.~M.
  2015.
\newblock Climate change sentiment on Twitter: An unsolicited public opinion
  poll.
\newblock \emph{PloS one}, 10(8): e0136092.

\bibitem[{Cogburn and Espinoza-Vasquez(2011)}]{cogburn2011networked}
Cogburn, D.~L.; and Espinoza-Vasquez, F.~K. 2011.
\newblock From networked nominee to networked nation: Examining the impact of
  Web 2.0 and social media on political participation and civic engagement in
  the 2008 Obama campaign.
\newblock \emph{Journal of political marketing}, 10(1-2): 189--213.

\bibitem[{Edelson et~al.(2019)Edelson, Sakhuja, Dey, and
  McCoy}]{edelson2019analysis}
Edelson, L.; Sakhuja, S.; Dey, R.; and McCoy, D. 2019.
\newblock An analysis of united states online political advertising
  transparency.
\newblock \emph{arXiv preprint arXiv:1902.04385}.

\bibitem[{Ennser-Jedenastik et~al.(2022)Ennser-Jedenastik, Gahn, Bodlos, and
  Haselmayer}]{ennser2022does}
Ennser-Jedenastik, L.; Gahn, C.; Bodlos, A.; and Haselmayer, M. 2022.
\newblock Does social media enhance party responsiveness? How user engagement
  shapes parties’ issue attention on Facebook.
\newblock \emph{Party Politics}, 28(3): 468--481.

\bibitem[{Facebook(2021{\natexlab{a}})}]{Facebook3}
Facebook. 2021{\natexlab{a}}.
\newblock Facebook Ad Library Report.
\newblock Data retrieved from Facebook Ad Library Report on 08.11.2021,
  \url{https://www.facebook.com/ads/library/report/}.

\bibitem[{Facebook(2021{\natexlab{b}})}]{facebook6}
Facebook. 2021{\natexlab{b}}.
\newblock Facebook Open Research \& Transparency (FORT) Ads Targeting Dataset.
\newblock
  \url{https://www.facebook.com/business/learn/lessons/verify-facebook-instagram-account}
  (visited on 26.07.2022).

\bibitem[{Facebook(2022)}]{Facebook5}
Facebook. 2022.
\newblock Ad Manager Metrics.
\newblock \url{https://www.facebook.com/business/help/264160060861852} (visited
  on 05.08.2022).

\bibitem[{Fowler et~al.(2021)Fowler, Franz, Martin, Peskowitz, and
  Ridout}]{fowler2021political}
Fowler, E.~F.; Franz, M.~M.; Martin, G.~J.; Peskowitz, Z.; and Ridout, T.~N.
  2021.
\newblock Political advertising online and offline.
\newblock \emph{American Political Science Review}, 115(1): 130--149.

\bibitem[{Husz{\'a}r et~al.(2022)Husz{\'a}r, Ktena, O’Brien, Belli,
  Schlaikjer, and Hardt}]{huszar2022algorithmic}
Husz{\'a}r, F.; Ktena, S.~I.; O’Brien, C.; Belli, L.; Schlaikjer, A.; and
  Hardt, M. 2022.
\newblock Algorithmic amplification of politics on Twitter.
\newblock \emph{Proceedings of the National Academy of Sciences}, 119(1):
  e2025334119.

\bibitem[{Hutto and Gilbert(2014)}]{hutto2014vader}
Hutto, C.; and Gilbert, E. 2014.
\newblock Vader: A parsimonious rule-based model for sentiment analysis of
  social media text.
\newblock In \emph{Proceedings of the international AAAI conference on web and
  social media}, volume~8, 216--225.

\bibitem[{IPCC(2021)}]{IPCC2021}
IPCC. 2021.
\newblock IPCC, 2021: Climate Change 2021.
\newblock Technical Report~6, Cambridge University Press, Cambrigde, UK.
\newblock \url{https://report.ipcc.ch/ar6wg1/index.html} (visited on
  23.11.2021).

\bibitem[{Jamison et~al.(2020)Jamison, Broniatowski, Dredze, Wood-Doughty,
  Khan, and Quinn}]{jamison2020vaccine}
Jamison, A.~M.; Broniatowski, D.~A.; Dredze, M.; Wood-Doughty, Z.; Khan, D.;
  and Quinn, S.~C. 2020.
\newblock Vaccine-related advertising in the Facebook Ad Archive.
\newblock \emph{Vaccine}, 38(3): 512--520.

\bibitem[{Kearney(2017)}]{kearney2017interpersonal}
Kearney, M.~W. 2017.
\newblock Interpersonal goals and political uses of Facebook.
\newblock \emph{Communication Research Reports}, 34(2): 106--114.

\bibitem[{Kl{\"u}ver and Sagarzazu(2016)}]{Klver2016SettingTA}
Kl{\"u}ver, H.; and Sagarzazu, I. 2016.
\newblock Setting the Agenda or Responding to Voters? Political Parties, Voters
  and Issue Attention.
\newblock \emph{West European Politics}, 39: 380 -- 398.

\bibitem[{Lewis et~al.(2019)Lewis, Liu, Goyal, Ghazvininejad, Mohamed, Levy,
  Stoyanov, and Zettlemoyer}]{lewis2019bart}
Lewis, M.; Liu, Y.; Goyal, N.; Ghazvininejad, M.; Mohamed, A.; Levy, O.;
  Stoyanov, V.; and Zettlemoyer, L. 2019.
\newblock {BART}: Denoising sequence-to-sequence pre-training for natural
  language generation, translation, and comprehension.
\newblock \emph{arXiv preprint arXiv:1910.13461}.

\bibitem[{Lutzke et~al.(2019)Lutzke, Drummond, Slovic, and
  {\'A}rvai}]{lutzke2019priming}
Lutzke, L.; Drummond, C.; Slovic, P.; and {\'A}rvai, J. 2019.
\newblock Priming critical thinking: Simple interventions limit the influence
  of fake news about climate change on Facebook.
\newblock \emph{Global environmental change}, 58: 101964.

\bibitem[{Matsa and Walker(2021)}]{matsa2021news}
Matsa, K.~E.; and Walker, M. 2021.
\newblock News Consumption Across Social Media in 2021.
\newblock Pew Research.

\bibitem[{Medina~Serrano, Papakyriakopoulos, and
  Hegelich(2020)}]{medina2020exploring}
Medina~Serrano, J.~C.; Papakyriakopoulos, O.; and Hegelich, S. 2020.
\newblock Exploring political ad libraries for online advertising transparency:
  lessons from Germany and the 2019 European elections.
\newblock In \emph{International conference on social media and society},
  111--121.

\bibitem[{Mejova and Kalimeri(2020)}]{mejova2020covid}
Mejova, Y.; and Kalimeri, K. 2020.
\newblock COVID-19 on Facebook ads: competing agendas around a public health
  crisis.
\newblock In \emph{Proceedings of the 3rd ACM SIGCAS Conference on Computing
  and Sustainable Societies}, 22--31.

\bibitem[{Meta(2021)}]{FB-social-issues-classified}
Meta. 2021.
\newblock Meta for Business.
\newblock \url{https://www.facebook.com/business/help/167836590566506} (visited
  on 29.06.2022).

\bibitem[{Metz(2021)}]{metz2021likes}
Metz, R. 2021.
\newblock Likes, anger emojis and RSVPs: The math behind Facebook’s News
  Feed—and how it backfired| CNN Business.

\bibitem[{Nott(2020)}]{nott2020political}
Nott, L. 2020.
\newblock Political advertising on social media platforms.
\newblock \emph{Human Rights Magazine}, 45(3).

\bibitem[{Owen(2019)}]{newmediapolitics}
Owen, D. 2019.
\newblock The New Media´s Role in Politics.
\newblock In \emph{Towards a New Enlightenment? A Transcendent Decade},
  347--365. BBVA OpenMind.

\bibitem[{Pearce et~al.(2019)Pearce, Niederer, {\"O}zkula, and
  S{\'a}nchez~Querub{\'\i}n}]{pearce2019social}
Pearce, W.; Niederer, S.; {\"O}zkula, S.~M.; and S{\'a}nchez~Querub{\'\i}n, N.
  2019.
\newblock The social media life of climate change: Platforms, publics, and
  future imaginaries.
\newblock \emph{Wiley interdisciplinary reviews: Climate change}, 10(2): e569.

\bibitem[{Reuning, Whitesell, and Hannah(2022)}]{reuning2022facebook}
Reuning, K.; Whitesell, A.; and Hannah, A.~L. 2022.
\newblock Facebook algorithm changes may have amplified local republican
  parties.
\newblock \emph{Research \& Politics}, 9(2): 20531680221103809.

\bibitem[{Sahly, Shao, and Kwon(2019)}]{sahly2019social}
Sahly, A.; Shao, C.; and Kwon, K.~H. 2019.
\newblock Social media for political campaigns: An examination of Trump’s and
  Clinton’s frame building and its effect on audience engagement.
\newblock \emph{Social Media+ Society}, 5(2): 2056305119855141.

\bibitem[{Sarewitz(2011)}]{sarewitz2011does}
Sarewitz, D. 2011.
\newblock Does climate change knowledge really matter?
\newblock \emph{Wiley Interdisciplinary Reviews: Climate Change}, 2(4):
  475--481.

\bibitem[{Statista(2021{\natexlab{a}})}]{statista-facebook-demographics}
Statista. 2021{\natexlab{a}}.
\newblock Distribution of Facebook users worldwide as of October 2021, by age
  and gender.
\newblock
  \url{https://www.statista.com/statistics/376128/facebook-global-user-age-distribution/}
  (visited on 30.11.2021).

\bibitem[{Statista(2021{\natexlab{b}})}]{statista-instagram-demographics}
Statista. 2021{\natexlab{b}}.
\newblock Distribution of Facebook users worldwide as of October 2021, by age
  and gender.
\newblock \url{
  https://www.statista.com/statistics/325587/instagram-global-age-group/}
  (visited on 27.07.2022).

\bibitem[{Statista(2021{\natexlab{c}})}]{statista-instagram-demographics_US}
Statista. 2021{\natexlab{c}}.
\newblock Distribution of Instagram users in the United States as of July 2021,
  by gender.
\newblock
  \url{https://www.statista.com/statistics/530498/instagram-users-in-the-us-by-gender/}
  (visited on 02.12.2021).

\bibitem[{Statista(2021{\natexlab{d}})}]{statista-facebook}
Statista. 2021{\natexlab{d}}.
\newblock Most popular social networks worldwide as of October 2021.
\newblock
  \url{https://www.statista.com/statistics/272014/global-social-networks-ranked-by-number-of-users/}
  (visited on 30.11.2021).

\bibitem[{Tarai et~al.(2015)Tarai, Kant, Finau, and
  Titifanue}]{tarai2015political}
Tarai, J.; Kant, R.; Finau, G.; and Titifanue, J. 2015.
\newblock Political social media campaigning in Fiji's 2014 elections.
\newblock \emph{Journal of Pacific Studies}, 35(3): 89--114.

\bibitem[{Tauberer et~al.(2021)Tauberer, Mill, Willis, and other
  contributors}]{Congress-legislators}
Tauberer, J.; Mill, E.; Willis, D.; and other contributors. 2021.
\newblock Congress-Legislators.
\newblock Data retrieved from GitHub repository on 12.10.2021,
  \url{https://github.com/unitedstates/congress-legislators}.

\bibitem[{{Tech For Campaigns}(2020)}]{tech4campaigns}
{Tech For Campaigns}. 2020.
\newblock 2020 Political Digital Advertising Report.
\newblock
  \url{https://www.techforcampaigns.org/impact/2020-political-digital-advertising-report}
  (visited on 10.11.2021).

\bibitem[{{TechCrunch}(2021)}]{techcrunch-facebook-api-start}
{TechCrunch}. 2021.
\newblock Facebook and Instagram launch US political ad labeling and archive.
\newblock
  \url{https://techcrunch.com/2018/05/24/facebook-political-ad-archive/}
  (visited on 07.12.2021).

\bibitem[{Thorson et~al.(2021)Thorson, Cotter, Medeiros, and
  Pak}]{thorson2021algorithmic}
Thorson, K.; Cotter, K.; Medeiros, M.; and Pak, C. 2021.
\newblock Algorithmic inference, political interest, and exposure to news and
  politics on Facebook.
\newblock \emph{Information, Communication \& Society}, 24(2): 183--200.

\bibitem[{United~Nations(2021)}]{UN_climate_change_biggest_threat}
United~Nations, U. 2021.
\newblock Climate Change ‘Biggest Threat Modern Humans Have Ever Faced’,
  World-Renowned Naturalist Tells Security Council, Calls for Greater Global
  Cooperation.
\newblock \url{https://www.un.org/press/en/2021/sc14445.doc.html} (visited on
  27.11.2021).

\bibitem[{Victor(2015)}]{victor2015climate}
Victor, D. 2015.
\newblock Climate change: Embed the social sciences in climate policy.
\newblock \emph{Nature}, 520(7545): 27--29.

\bibitem[{Vu et~al.(2021)Vu, Blomberg, Seo, Liu, Shayesteh, and
  Do}]{vu2021social}
Vu, H.~T.; Blomberg, M.; Seo, H.; Liu, Y.; Shayesteh, F.; and Do, H.~V. 2021.
\newblock Social media and environmental activism: Framing climate change on
  Facebook by global NGOs.
\newblock \emph{Science communication}, 43(1): 91--115.

\bibitem[{Watts et~al.(2018)Watts, Amann, Arnell, Ayeb-Karlsson, Belesova,
  Berry, Bouley, Boykoff, Byass, Cai et~al.}]{climate_change_health}
Watts, N.; Amann, M.; Arnell, N.; Ayeb-Karlsson, S.; Belesova, K.; Berry, H.;
  Bouley, T.; Boykoff, M.; Byass, P.; Cai, W.; et~al. 2018.
\newblock The 2018 report of the Lancet Countdown on health and climate change:
  shaping the health of nations for centuries to come.
\newblock \emph{The Lancet}, 392(10163): 2479--2514.

\bibitem[{Williams et~al.(2015)Williams, McMurray, Kurz, and
  Lambert}]{williams2015network}
Williams, H.~T.; McMurray, J.~R.; Kurz, T.; and Lambert, F.~H. 2015.
\newblock Network analysis reveals open forums and echo chambers in social
  media discussions of climate change.
\newblock \emph{Global environmental change}, 32: 126--138.

\bibitem[{Yu et~al.(2021)Yu, Margolin, Fownes, Eiseman, Chatrchyan, and
  Allred}]{yu2021tweeting}
Yu, C.; Margolin, D.~B.; Fownes, J.~R.; Eiseman, D.~L.; Chatrchyan, A.~M.; and
  Allred, S.~B. 2021.
\newblock Tweeting about climate: Which politicians speak up and what do they
  speak up about?
\newblock \emph{Social Media+ Society}, 7(3): 20563051211033815.

\end{thebibliography}

\end{document}